\documentclass[prb,preprint,superscriptaddress,longbibliography]{revtex4-1}
\usepackage{graphicx}
\usepackage{hyperref}
\usepackage{amsmath}
\usepackage{amssymb}
\usepackage[usenames,dvipsnames]{color}

\begin{document}

\title{Tuneable topological domain wall states in engineered atomic chains}

\author{Md Nurul Huda}
\affiliation{Department of Applied Physics, Aalto University School of Science, PO Box 15100, 00076 Aalto, Finland}

\author{Shawulienu Kezilebieke}
\affiliation{Department of Applied Physics, Aalto University School of Science, PO Box 15100, 00076 Aalto, Finland}

\author{Teemu Ojanen}
\affiliation{Department of Applied Physics, Aalto University School of Science, PO Box 15100, 00076 Aalto, Finland}
\affiliation{Computational Physics Laboratory, Physics Unit, Faculty of Engineering and Natural Sciences, Tampere University, PO Box 692, FI-33014 Tampere, Finland}

\author{Robert Drost}
\affiliation{Max-Planck-Institute for Solid State Research, Nanoscale Science Department, Heisenbergstrasse 1, D-70569 Stuttgart, Germany}

\author{Peter Liljeroth}
\email{Email: peter.liljeroth@aalto.fi}
\affiliation{Department of Applied Physics, Aalto University School of Science, PO Box 15100, 00076 Aalto, Finland}

\date{\today}


\begin{abstract}

Topological modes in one- and two-dimensional systems have been proposed for numerous applications utilizing their exotic electronic responses. The 1D, zero-energy, topologically protected end modes can be realized in structures implementing the Su-Schrieffer-Heeger (SSH) model. While the edge modes in the SSH model are at exactly the mid-gap energy, other paradigmatic 1D models such as trimer and coupled dimer chains have non-zero energy boundary states. However, these structures have not been realized in an atomically tuneable system that would allow explicit control of the edge modes. Here, we demonstrate atomically controlled trimer and coupled dimer chains realized using chlorine vacancies in the c$(2\times2)$ adsorption layer on Cu(100). This system allows wide tuneability of the domain wall modes that we experimentally demonstrate using low-temperature scanning tunneling microscopy (STM). 

\end{abstract}

\maketitle

\section*{Introduction}
The essential physics of quantum materials can be often captured using tight-binding (TB) models describing hopping between localized electronic orbitals. The argument can also be reversed: given sufficient control, it is possible to realize experimentally artificial materials by a suitable arrangement of coupled ``sites''. This can be achieved using atom manipulation by the tip of a scanning tunneling microscope (STM), which allows placing each atom individually into a well-defined, pre-determined position \cite{Eigler:1990atom_manipulation,Crommie1993,Manoharan2000,Gomes2012,celotta2014,Folsch:2014NatNano,Kalff:NatNano2016,Drost:NatPhys2017,Girovsky:SciPost2017,Slot:2017NatPhys}. It is thus possible to build designer quantum materials with ultimate control over their electronic structure through atomic assemblies. This has been used to demonstrate formation of quantum confined one- or two-dimensional electronic systems, and artificial lattices with e.g. honeycomb and Lieb symmetries \cite{Crommie1993,Folsch2004_PRL,Folsch:2014NatNano,Repp2005,Schuler2015,Gomes2012,Paavilainen2016,Drost:NatPhys2017,Slot:2017NatPhys,Girovsky:SciPost2017,Slot2018}.  
\begin{figure}[!b]
\centering
\includegraphics{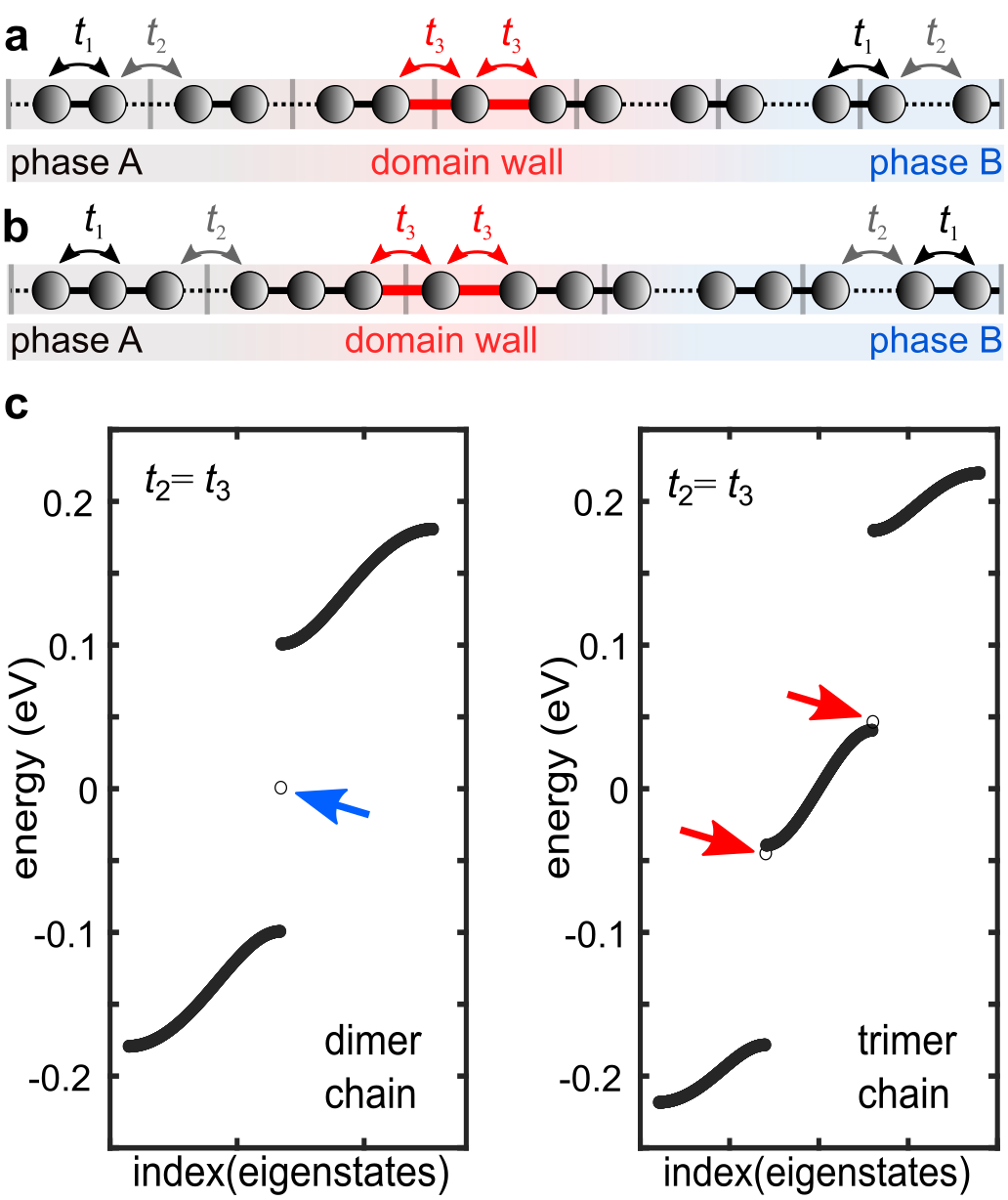}
\caption{Domain walls in dimer and trimer chains. \textbf{a,b} Schematic of dimer (a) and trimer (b) chain with a domain wall. \textbf{c} Calculated energy spectra of finite dimer and trimer chains hosting a domain wall with $t_1=0.14$ eV, $t_2=0.04$ eV \cite{Drost:NatPhys2017} calculated with 300 unit cells on both sides of the domain wall. While the in-gap state in a dimer chain is pinned to mid-gap, domain wall in a trimer chain results tunable in-gap states depending on the hopping $t_3$.}
\label{fig:one}
\end{figure}

Topological modes in one- and two-dimensional systems have been proposed for numerous applications utilizing their exotic electronic responses \cite{Hasan2010,mourik2012,Nadj-Perge2014,Ruby2015_prl_2,Wiesendanger2018_Majorana,Cheon2015,Yeom2017NatPhys,Louie2017_GNR,Sato2017_review}. Atomic manipulation can naturally also be applied to topological materials and systems with interface and edge states can be constructed \cite{Drost:NatPhys2017}. In one-dimensional (1D) systems, the dimer chain (Fig.~\ref{fig:one}a, Su-Schrieffer-Heeger model realized in e.g.~polyacetylene) is the prototypical example of a 1D topological material \cite{heeger}. The system comprises of dimers with a strong hopping between the sites ($t_1$), and weaker hopping between the dimers ($t_2$). Depending on the choice of the unit cell (see Fig.~1a), the chain can exist in two, topologically distinct, ground states. The bulk band structure is unchanged when comparing two infinite chains displaced by a fraction of a unit cell. However, merging two such displaced gapped chains gives rise to a domain wall and an in-gap state spatially localized in the vicinity of the domain wall \cite{jackiw1976,heeger,Drost:NatPhys2017,Kane2013}. These zero-energy modes have been experimentally implemented in atomic-scale solid-state structures and in ultra-cold atomic gases \cite{Drost:NatPhys2017,Meier2016_UGC}, and analogous systems have been realized in graphene nanoribbons \cite{Louie2017_GNR,Fasel2018,Crommie2018}. This state is protected in the sense that it is impossible to get rid of it without closing the bulk energy gap and it will always occur when joining two dimer chains with displaced unit cells. At half-filling, the existence of this state can be understood in the form of a soliton state with a polarization charge $\pm e/2$ on the boundary of the two ground states of dimer chain. For this simplest realization of a dimer chain, the in-gap state is exactly at the mid-gap energy.

The existence of a domain wall state at exactly mid-gap energy does not necessarily occur for 1D chains consisting of sites with different on-site energies or more complicated unit cells with three or more atoms \cite{Schrieffer1981_PRL,Rice:1982soliton,Jackiw:1983soliton,Kivelson:1983soliton,Cheon2015,Yeom2017NatPhys,Martinez2018_trimer}. While the states still necessarily occur at the domain wall, their energies within the gap can be tuned. A seminal contribution on self-assembled indium atomic wires on Si(111) demonstrated the formation of coupled dimer chains with four topological distinct bulk phases \cite{Cheon2015}. Cheon et al. identified different kinds of interfaces between the phases and showed that the interface states in these double Peierls chains can be understood as topological chiral solitons \cite{Cheon2015,Yeom2017NatPhys}.

The self-assembled chains can be influenced by defects and other imperfections, and they have a stochastic distribution of domain walls with different structures and properties \cite{Yeom2012_PRL,Cheon2015,Lee2015_In-Si,Lee2017,Lee2019_PRL}. This can be overcome using atomic manipulation that allows the realization of perfect, defect-free structures with predetermined types of domain walls. We use atomic manipulation with a low-temperature STM to realize paradigmatic 1D model systems of trimer and coupled dimer chains with precisely controlled domain wall structures. Here, we use the localized electronic states hosted in chlorine vacancies in the chlorine c$(2\times2)$ structure on Cu(100) as building blocks, which have recently be used to create artificial systems with designer electronic structures \cite{Kalff:NatNano2016,Drost:NatPhys2017,Girovsky:SciPost2017}. We focus on linear systems where the unit cell is more complicated (trimer and coupled dimer chains) and demonstrate the formation of tuneable interface states between different ground states of the system. The domain walls between different trimer and coupled dimer states can in principle be used to prepare localized and well-defined fractional charges and to manipulate them. In the future, these modes may find applications in exotic quantum devices with atomically well-defined geometries \cite{heeger,Schrieffer1981_PRL,Thouless1983_PRB,Troyer2013_PRL}.

\section*{Results and Discussion}

Fig.~\ref{fig:one}b shows a schematic of a trimer chain. Similarly to dimer chains, joining sections of trimer chains with different unit cells necessarily results in the formation of a domain wall that cannot be removed from the system by a local perturbation. Perfectly trimerized chains (intratrimer hopping $t_1$, and intertrimer hopping $t_2$) have three distinct topological phases and an electronic structure with three separate bands. In a system containing a domain wall, localized states appear in the band gaps\cite{Schrieffer1981_PRL,Guo2014_PhysLett}. Domain walls host a fractional charge of $\pm 1/3e$ or $\pm2/3e$ depending on domain wall type and the chemical potential of the system. These charges appear due to the mismatch of the unit cells on the opposite sides of the domain wall and cannot be removed without changing the bulk structure of the chain \cite{Schrieffer1981_PRL}. However, the energy position of the state associated with domain wall is not fixed and can be moved within the gap; the states are movable, but irremovable. The electronic structure based on TB calculations in finite dimer and trimer chains are illustrated in Fig.~\ref{fig:one}c. The TB parameters correspond to the experimental values of the chlorine vacancy system \cite{Drost:NatPhys2017} (details of the TB calculations are given in the Supporting Information (SI)). Analyzing the nature and the number of the localized states, it is clear that they arise from bonding and antibonding combinations between the states on the domain wall site and the nearest trimer units. 

\begin{figure}[!t]
\centering
\includegraphics{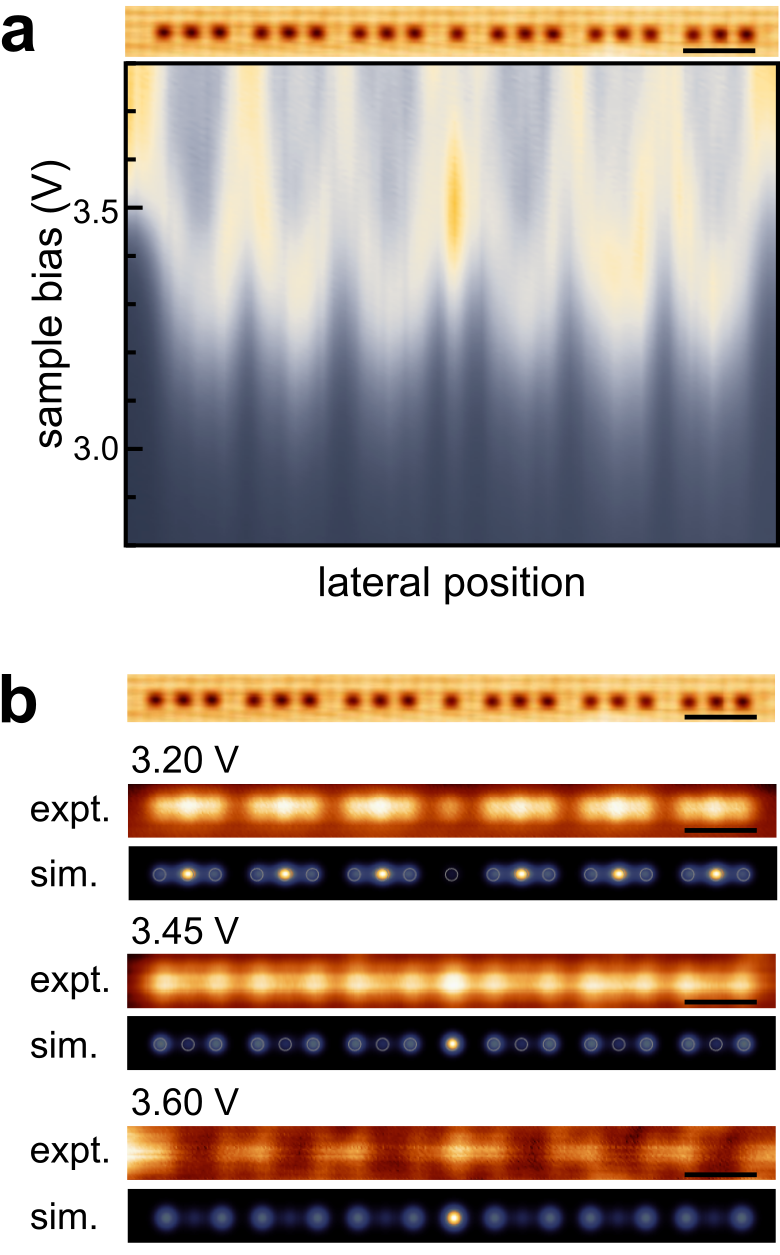}
\caption{Experimental realization of a trimer chain. \textbf{a} Stacked contour plot of d$I$/d$V$ spectra taken along a trimer chain. The chain topography is shown on top for reference (bias 0.5 V, current 1 nA). \textbf{b} STM topography of trimer  (top), experimental d$I$/d$V$ and simulated LDOS maps at the biases indicated in the figure. d$I$/d$V$ maps are recorded in the constant-height mode (determined by set-point current of 1 nA). Theoretical plots are based on a tight-binding model with $t_1=0.14$ eV, $t_2=t_3=0.04$ eV. Scale bars, 2 nm.}
\label{fig:two}
\end{figure}
Fig.~\ref{fig:two}a shows a realization of a trimer chain with a domain wall in the chlorine vacancy system introduced in Refs.~\cite{Kalff:NatNano2016,Drost:NatPhys2017,Girovsky:SciPost2017}. Sample preparation and the details of the STM experiments are described in the Methods section and the SI. In this experimental system, it is difficult to access the higher energy gap of the trimer chain as it is close to the conduction band of the chlorine layer \cite{Drost:NatPhys2017}. In addition to the STM topography, Fig.~\ref{fig:two}a shows d$I$/d$V$ spectra measured along the trimer chain. The localized states at the domain wall are clearly visible at the bias of around 3.5 V. The localized state can be clearly visualized also in the spatially-resolved d$I$/d$V$ maps, where low energy maps show intensity over the whole chain and the domain wall states are visible at the biases corresponding to energies close to the on-site energy (in line with TB predictions). Data for another type of a domain wall in the trimer chain is shown in the SI.

\begin{figure*}[!th]
\centering
\includegraphics[width=0.6\textwidth]{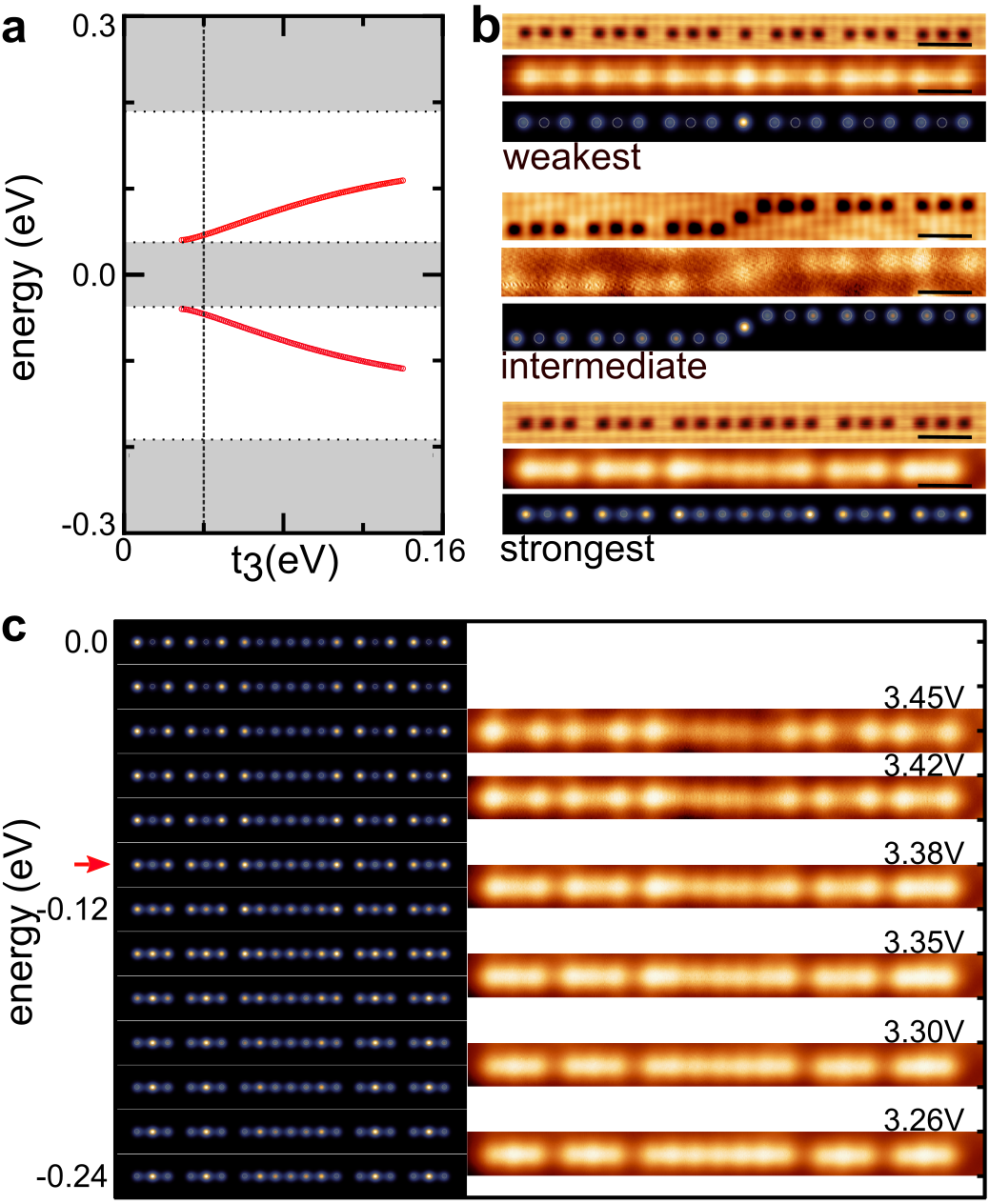}
\caption{Tuning the energies of the domain wall states. \textbf{a} Energies (w.r.t.~the on-site energy) of the trimer chain in-gap states depending on the hopping $t_3$ calculated with a tight-binding model with $t_1=0.14$ eV, $t_2=0.04$ eV. The shaded areas indicate the three bands of the bulk trimer chain and the dashed vertical line marks $t_3=t_2$. \textbf{b} Experimental realizations of modulating the hopping $t_3$ into the domain wall site ($t_3=0.04;~0.07;~0.14$ eV). The panels are labelled with the strength of the hopping. In each case, we show STM topography (top, set-point 0.5 V/1 nA), d$I$/d$V$ map acquired at a bias close to the in-gap state (3.45, 3,48, and 3.38 V, respectively) and the corresponding simulated LDOS map. Scale bars, 2 nm. \textbf{c} Comparison between the simulated LDOS maps in the case $t_3=0.14$ eV and the experimental constant-height d$I$/d$V$ maps as a function of energy. The red arrow marks the energy of the in-gap state. }
\label{fig:three}
\end{figure*}

With atomic level control, we can move away from the perfect dimerization or trimerization and tune the hoppings between the bulk chains and the domain wall. In the case of a dimer chain, this is expected to have no effect on the energy spectrum as the domain wall state is at zero energy due to symmetry reasons. However, in the case of a trimer chain, we expect that we can move the states within the band gaps. This is illustrated in Fig.~\mbox{\ref{fig:three}}, where panel a shows the calculated energies of the domain wall state as the hopping $t_3$ onto the domain wall site is tuned. The energies are given w.r.t.~the on-site energy. The states remain in the gaps as long as the hopping $t_3$ is not much smaller than the weaker hopping $t_2$ in the chain. By adjusting the nearest neighbour distances, we experimentally tune the hopping $t_3$ from 0.04\,eV (weakest) to 0.14\,eV (strongest) (see Fig.~\mbox{\ref{fig:three}}b).

Extracting the energies of the domain wall states directly from the experimental d$I$/d$V$ is difficult due to the energy broadening of the spectra. In addition, the higher lying states coincide with the conduction band of the chlorine layer and hence cannot be reliably detected (see SI for details). Instead, we can demonstrate that the experimental results are consistent with the simulations without freely adjustable parameters. The on-site energy, the energy broadening, and the spatial shape of the square of the wavefunction can be extracted from experiments on single vacancies. We get $3.49\pm0.01$ V as the on-site energy, the lineshape is a lorentzian with half-width at half-maximum $\Delta = \pm0.01$ eV, and the spatial shape is given by a gaussian with a full-width at half-maximum of $\Gamma = 0.71a$ ($a$ is the lattice constant of the c($2\times2$)-Cl structure). The values of the hoppings can be extracted from experiments on isolated dimers, which yields the values of $t_3=$~0.04, 0.07, or 0.14 eV for the three structures shown in Fig.~\mbox{\ref{fig:three}}b. 

Using these values, we can simulate a series of LDOS maps and compare them to the experimental ones. This is illustrated in Fig.~\mbox{\ref{fig:three}}c and it can be seen that there is an excellent correspondence between the simulations and experimental constant-height d$I$/d$V$ maps (comparison for the weakest coupling is shown in the SI). Similarly for the other domain wall structures, we can match the LDOS maps at the bias corresponding to the domain wall state to the simulated LDOS maps as shown in Fig.~\mbox{\ref{fig:three}}a. The domain walls between trimer chains shifted by one-third (two-thirds) of a unit cell support $e/3$ ($2e/3$) charge per spin when the lowest band is filled \cite{heeger,Schrieffer1981_PRL}. The possible charge states at a domain wall are integer multiples of $e/3$, being insensitive to the precise domain wall geometry and in this sense a topological property. Despite the long history of the theoretical studies of trimer chains and their domain wall states, the present work is the first step towards realizing their exotic properties in solid state devices.

\begin{figure}[!th]
\centering
\includegraphics{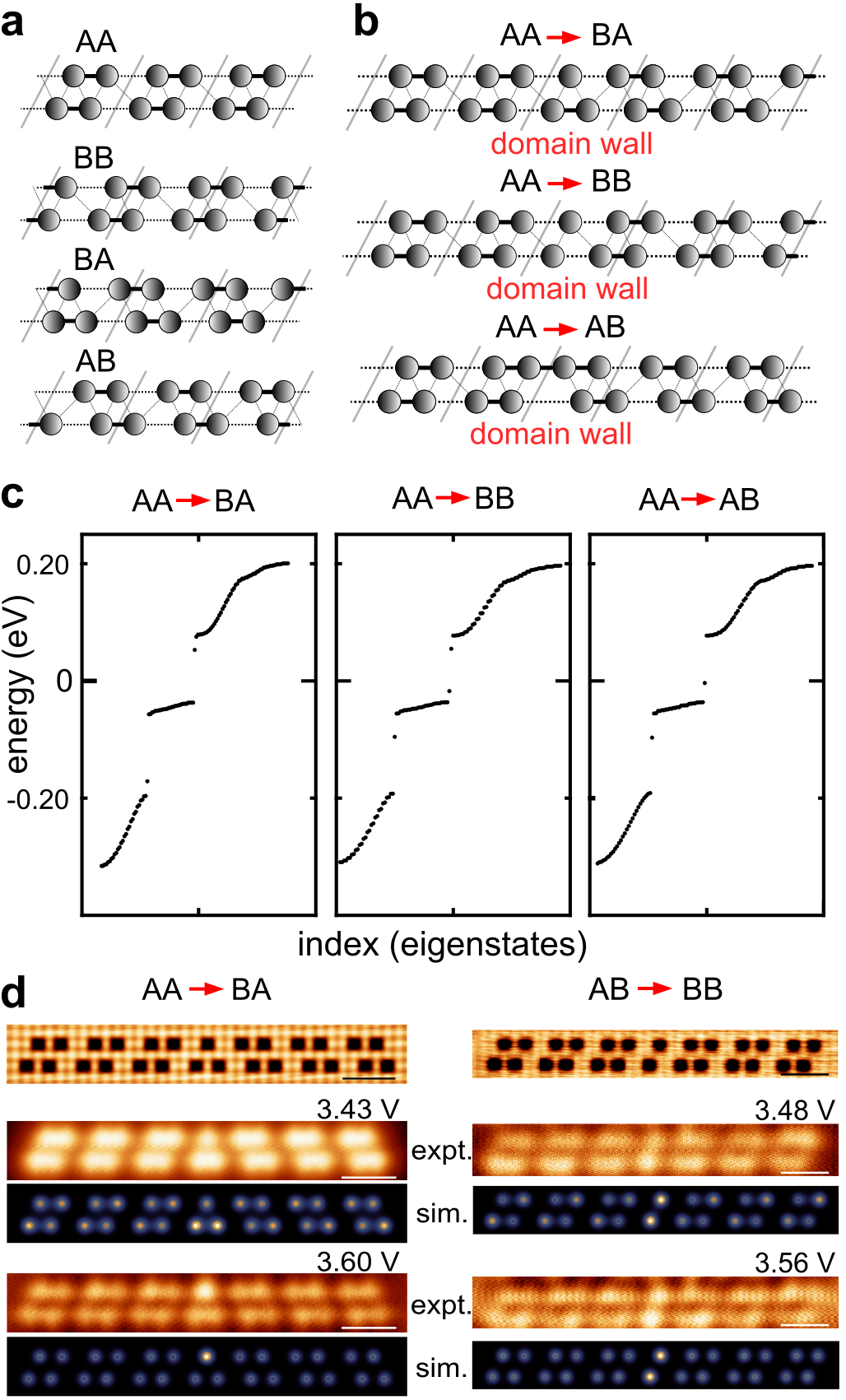}
\caption{Domain wall states in coupled dimer chains. \textbf{a} The four ground states of a coupled dimer chains. \textbf{b} different kinds of domain walls that we have constructed. \textbf{c} The energy diagrams of the domain walls shown in panel (a) (calculated with 7 unit cells on each side of the domain wall). \textbf{d} Experimental realizations of the $AA\rightarrow BA$ and $AA\rightarrow BB$ domain walls. Scale bars, 2 nm.}
\label{fig:four}
\end{figure}

In addition to the tuneable energy of the domain wall states in trimer chains, we consider tuneable domain wall electronic states in coupled dimer chains. This type of systems were demonstrated for the first time in the self-assembled coupled indium dimer chains on silicon \cite{Cheon2015,Yeom2017NatPhys,Lee2019_PRL}. Atomic level control allows us to go further and fabricate arbitrary domain walls in coupled dimer chains. Our artificial system readily gives access to domain walls between any of the four different unit cell geometries (Fig.~\ref{fig:four}a). We have realized all the structures shown in Fig.~\ref{fig:four}b and characterized the domain wall states using d$I$/d$V$ spectroscopy and mapping (see SI for additional results). The calculated band structures for the domain walls in in Fig.~\ref{fig:four}b are shown in Fig.~\ref{fig:four}c. The bulk chain has in principle four bands, but using the hoppings corresponding to our experimental system, the two higher energy bands overlap. 

Despite that chiral symmetry protecting topological phases and midgap domain wall states in a single dimer chain is broken in coupled chains, the existence of the domain wall states in the coupled chains have  a hidden topological origin as pointed out in Ref.~\citenum{Cheon2015}. 
The different states of the double dimer chain can be  expressed in terms of phases $A$ and $B$ (Fig.~\ref{fig:one}a) to denote the dimerization in each phase, i.e.~the coupled dimer chains can be categorized as $AA$, $AB$, $BA$, or $BB$, corresponding to both chains being in phase $A$, one being in phase $A$ and the other in phase $B$ etc. The domain wall states can then be classified based on whether they shift the dimerization in only one (e.g., $AA\rightarrow AB$ or $AA\rightarrow BA$) or both chains (e.g., $AA\rightarrow BB$). As discussed in Ref.~\citenum{Cheon2015}, the topological origin of the domain wall states is revealed by considering a formal pumping processes between different dimerizations that will transform the system cyclically through configurations ($AA\rightarrow AB\rightarrow BB\rightarrow BA\rightarrow AA$) \cite{Cheon2015,Yeom2017NatPhys}. The domain wall states can then be interpreted as snap shots of quantum-Hall type chiral edge modes in the formal 2D system of dimerization variable and physical chain. Since the energies of the chiral modes traverse the energy gap depending on the dimerization pattern, their energies are tuneable by the specific geometry of the domain wall. Nevertheless, the domain wall states cannot be removed from the gap by small perturbations. In our case, the coupling between the chains is strong and the spectrum does not closely resemble a decoupled chain. Nevertheless, we can construct analogous domain walls that act differently on the dimerizations of the two chains. Fig.~\ref{fig:four}d shows examples of domain walls that either shift the dimerization in only a single chain (left) or in both chains (right). 

The energies of the domain wall states are not equal in the different cases and they do depend on the values of the hoppings and the domain wall type. To get a general idea on the behaviour, we can consider states arising from the domain wall segment and how it couples to the rest of the chain. For example, for the structure $AA\rightarrow BA$, the domain wall itself consists of three lattice sites. There are three states associated with this (see SI for a schematic of the energy levels as a function of the interchain coupling ($t’$)). For the experimental value of the interchain coupling, the highest energy domain wall state overlaps with a bulk band and hybridizes with it and consequently, the system has two states within the gaps. Furthermore, the energies of these states are influenced on how they are coupled with the bulk chains (similarly to the case of the trimer chain).

In the case of the $AA\rightarrow BB$ domain wall, there are three in-gap states. The two sites forming the domain wall contribute two states that form bonding and antibonding combinations (the lowest and highest in energy, respectively). The third in-gap state is a hybrid between the domain wall and the middle band. We show the calculated wavefunctions of these states in the SI. Analogous arguments can be made for the rest of the possible domain wall structures. Again, the experimental results are fully in-line with the TB predictions, indicating that the simple model can be used to design more complicated structures.


In conclusion, we have demonstrated engineering domain wall states in artificial structures fabricated with atomic level control. Trimer chains allow fabricating domain walls where energy level positions can be tuned through the coupling between the bulk chain and the domain wall site. More complicated structures that can be realized in coupled dimer chains allow domain wall states with additional degrees of freedom (chirality). In the future, extending the atomic manipulation using automated schemes \cite{Kalff:NatNano2016,Girovsky:SciPost2017} will make it possible to test ideas on using domain wall states with fractional charges (depending on the chemical potential of the system), topological charge pumping and other exotic quantum devices.

\section*{Methods}

Sample preparation\\ 
All sample preparations and experiments were carried out in an ultrahigh vacuum system with a base pressure of $\sim$10$^{-10}$ mbar. The (100)-terminated copper single crystal was cleaned by repeated cycles of Ne$^{+}$ sputtering at 1.5\,kV, annealing to 600\,$^{\circ}$C. To prepare the chloride structure, anhydrous CuCl$_2$ was deposited from an effusion cell held at 300$^{\circ}$C onto the warm crystal ($T\approx$ 150 - 200$^{\circ}$C) for 180 seconds. The sample was held at the same temperature for 10 minutes following the deposition.

STM experiments\\
After the preparation, the sample was inserted into the low-temperature STM (Unisoku USM-1300) and all subsequent experiments were performed at $T=4.2$ K. STM images were taken in the constant current mode. d$I$/d$V$ spectra were recorded by standard lock-in detection while sweeping the sample bias in an open feedback loop configuration, with a peak-to-peak bias modulation of 20~mV at a frequency of 709~Hz. Line spectra were acquired in constant height; the feedback loop was not closed at any point between the acquisition of the first and last spectra. Manipulation of the chlorine vacancies was carried out as described previously \cite{Kalff:NatNano2016,Drost:NatPhys2017}. The tip was placed above a Cl atom adjacent to a vacancy site at 0.5~V bias voltage and the current was increased to 1 to 2~$\mu$A with the feedback circuit engaged. The tip was then dragged towards the vacancy site at a speed of up to 250~pm/s until a sharp jump in the $z$-position of the tip was observed. This procedure lead to the Cl atom and the vacancy site exchanging positions with high fidelity.

\section*{Acknowledgements}
This research made use of the Aalto Nanomicroscopy Center (Aalto NMC) facilities and was supported by the European Research Council (ERC-2017-AdG no.~788185 ``Artificial Designer Materials''), Academy of Finland (Academy professor funding no.~318995 and 320555, and Academy postdoctoral researcher no.~309975), and the Aalto University Centre for Quantum Engineering (Aalto CQE).

\bibliography{bibliography}

\end{document}


\title{Supplementary Information:\\Tuneable topological domain wall states in engineered atomic chains}

\author{Md Nurul Huda}
\affiliation{Department of Applied Physics, Aalto University School of Science, PO Box 15100, 00076 Aalto, Finland}

\author{Shawulienu Kezilebieke}
\affiliation{Department of Applied Physics, Aalto University School of Science, PO Box 15100, 00076 Aalto, Finland}

\author{Teemu Ojanen}
\affiliation{Department of Applied Physics, Aalto University School of Science, PO Box 15100, 00076 Aalto, Finland}
\affiliation{Computational Physics Laboratory, Physics Unit, Faculty of Engineering and Natural Sciences, Tampere University, PO Box 692, FI-33014 Tampere, Finland}

\author{Robert Drost}
\affiliation{Max-Planck-Institute for Solid State Research, Nanoscale Science Department, Heisenbergstrasse 1, D-70569 Stuttgart, Germany}

\author{Peter Liljeroth}
\email{Email: peter.liljeroth@aalto.fi}
\affiliation{Department of Applied Physics, Aalto University School of Science, PO Box 15100, 00076 Aalto, Finland}

\maketitle
\renewcommand\thefigure{S\arabic{figure}} 
\renewcommand{\thesection}{\arabic{section}}

\section{Tight-binding calculations}

\begin{figure}[!h]
\centering
\includegraphics [width=.5\columnwidth] {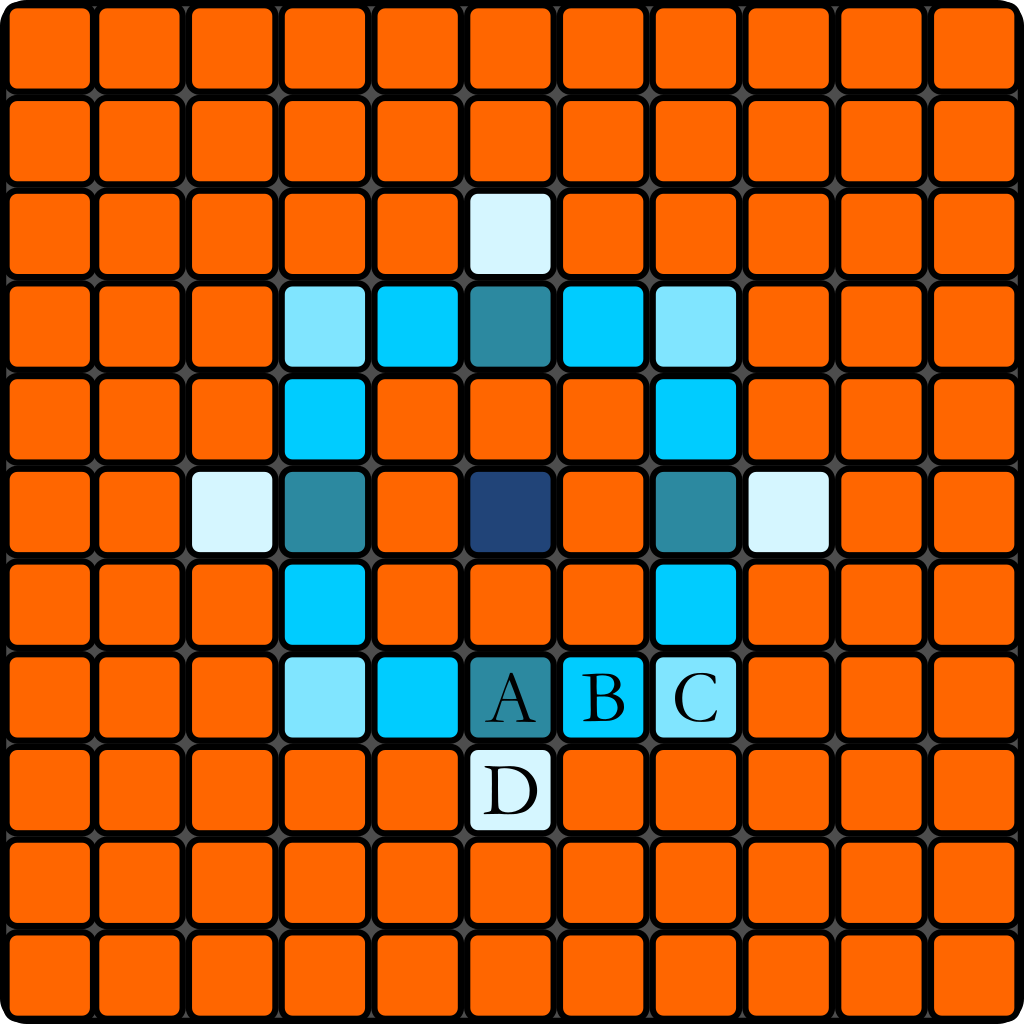}
\caption{Sketch of the c($2\times2$)-Cl adsorption structure. Letters A through D mark the nearest through fourth nearest-neighbor sites considered in the model with respect to the center site (note that there must always be at least one chlorine atom separating vacancy sites in the experimental structures). Identical colors denote identical coupling.}
\label{fig:SI_Fig1}
\end{figure}

The tight binding model is parametrised on the basis of our earlier work\cite{Drost:NatPhys2017}. The numerical values of the hopping amplitudes are (see also Fig.~\ref{fig:SI_Fig1}):

\begin{equation}
    \begin{array}{cc}
    t_{NN} = -0.14\ \textrm{eV} & t_{NNN} = -0.07\ \textrm{eV}\\
    t_{3NN} = -0.05\ \textrm{eV} & t_{4NN} = -0.04\ \textrm{eV}
    \end{array} \nonumber
\end{equation}

Atomic positions can be entered into a graphical user interface coded in Matlab representing the chlorine adsorption structure on Cu(100) to simulate structures of interest. The Matlab code implements the tight binding Hamiltonian 
\begin{equation}
    \mathscr{H}=-\sum_{ij} t_{ij} \hat{c}_i^{\dagger} \hat{c}_{j}+ h.c.
\end{equation}
where the summation runs over all pairs of sites and the hopping amplitudes are non-zero for the terms shown in Fig.~\ref{fig:SI_Fig1}. The on-site energy is set to $3.49\pm0.01$ V (value obtained experimentally from spectroscopy on a single vacancy). The resulting hopping matrix is diagonalised to obtain the eigenvectors and values. 

Simulated local density of states (LDOS) maps at an energy $\epsilon_0$ are drawn according to:
\begin{equation}
    \rho(x,y,\epsilon_0) = \sum_j \left[ \sum_{i = 1}^{N} v_i(\epsilon_j) \mathrm{e}^{-((x-x_i)^2 - (y-y_i)^2)/(2\Gamma^2)} \right]^2 \frac{1}{\pi}\frac{\Delta}{(\epsilon_j-\epsilon_0)^2+\Delta^2}
\end{equation}
where $j$ is the eigenstate index, $N$ is the total number of vacancy sites in the system,  $\epsilon_j$ and $v(\epsilon_j)$ are the energy and eigenvector corresponding to the $j^{th}$ eigenvalue, $(x_i,\ y_i)$ the position of the $i^{th}$ vacancy, $\Gamma$ is a phenomenological spatial broadening, and $\Delta$ the energy broadening.

To obtain a simulated LDOS contour, for each eigenvalue, a two-dimensional Gaussian contour is placed at each vacancy site and scaled by the eigenvector entry corresponding to that site at a given eigenvalue. The complete map for that eigenvalue is squared and weighted with a lorentzian  according to difference between its eigenenergy and the energy of interest. The complete map is then the sum over all these constituents.

Good agreement with the experimental data is reached for $\Delta = 0.18\pm0.01$ eV and $\Gamma = 0.71\ a$, where $a$ is the lattice constant of the c($2\times2$)-Cl structure.

\newpage
\section{Convergence of the trimer chain states as a function of the chain length}
\begin{figure}[!h]
\centering
\includegraphics [width=.9\columnwidth] {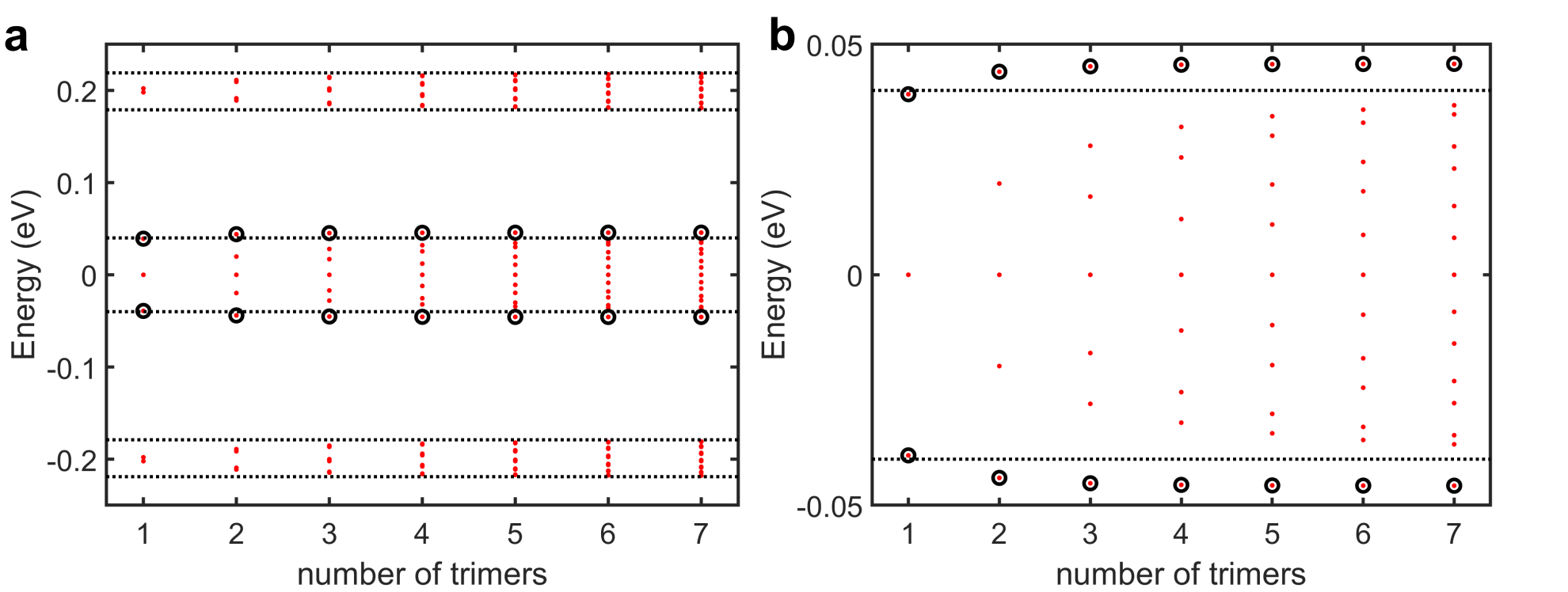}
\caption{\textbf{a,b} Simulated convergence of the trimer chain electronic structure. The chain consists of a number of trimer units ($x$-axis) on both sides of the domain wall with the experimental values of the hopping. The full energy range is shown in panel a and panel b is a zoom-in. Solid small red dots represent states and black circles indicate the domain wall states.}
\label{fig:S_conv}
\end{figure}

We checked for the effect of finite size effects using TB calculations and they are not very significant in this case and below (Fig.~\mbox{\ref{fig:S_conv}}) we show simulated results for the convergence of the band structure and the domain wall states. Given the broadening, we reach ``bulk behavior'' in already quite short trimer chains. This is especially true concerning the energy position of the domain wall states.

\newpage
\section{z-drift correction of d$I$/d$V$ spectra taken on a trimer chain} 
\begin{figure}[!h]
\centering
\includegraphics [width=.8\columnwidth] {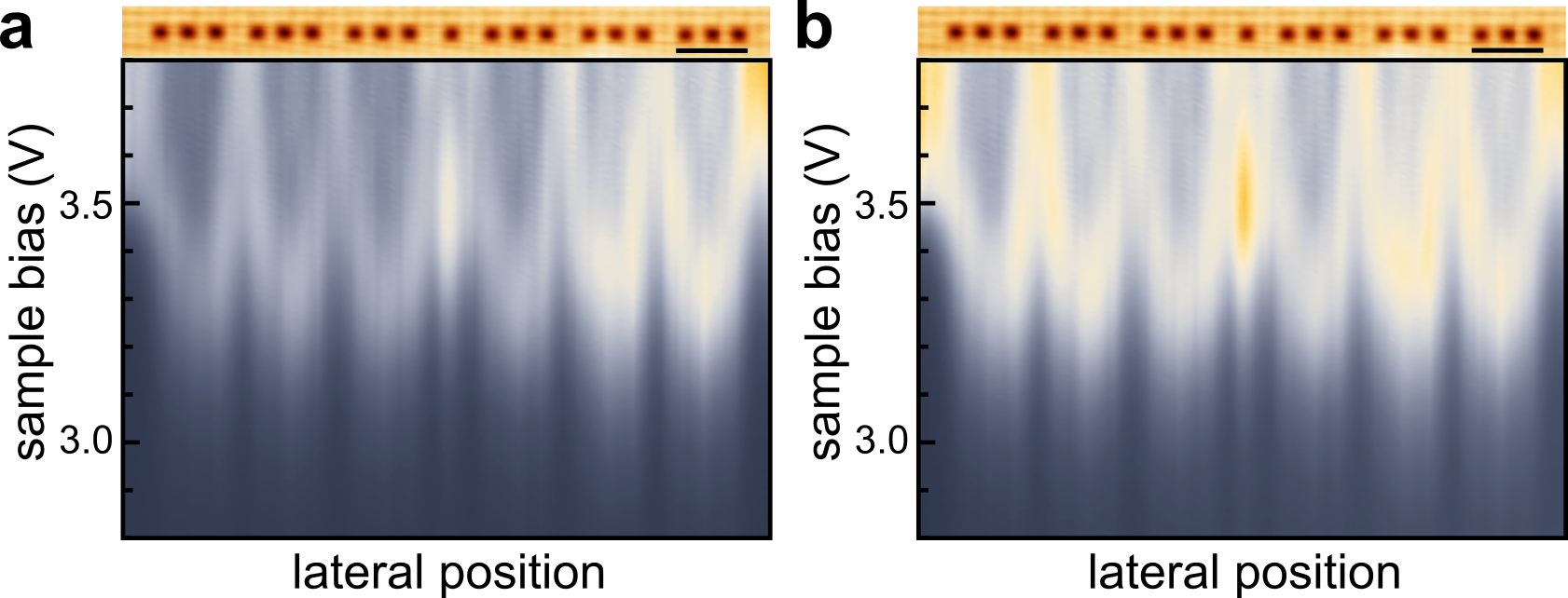}
\caption{Stacked d$I$/d$V$ spectra recorded over a trimer chain. \textbf{a} Raw experimental data. \textbf{b} Experimental $z$-drift removed by assuming constant drift. (STM images taken at 1 nA / 0.5 V, scale bar 2 nm).}
\label{fig:s_4}
\end{figure}

The experiment shown in Fig.~2a of the main paper was carried out in the constant-height mode (feedback loop was disconnected for the entire duration of taking the spectra) and there was some residual $z$-drift causing the tip-sample distance to increase slightly during the experiment. This was corrected in a post processing step assuming constant drift. Fig.~\ref{fig:s_4}a shows the original raw experimental data.

\newpage
\section{Energy broadening and the states above the on-site energy in the trimer chains}
\begin{figure}[!h]
\centering
\includegraphics [width=.8\columnwidth] {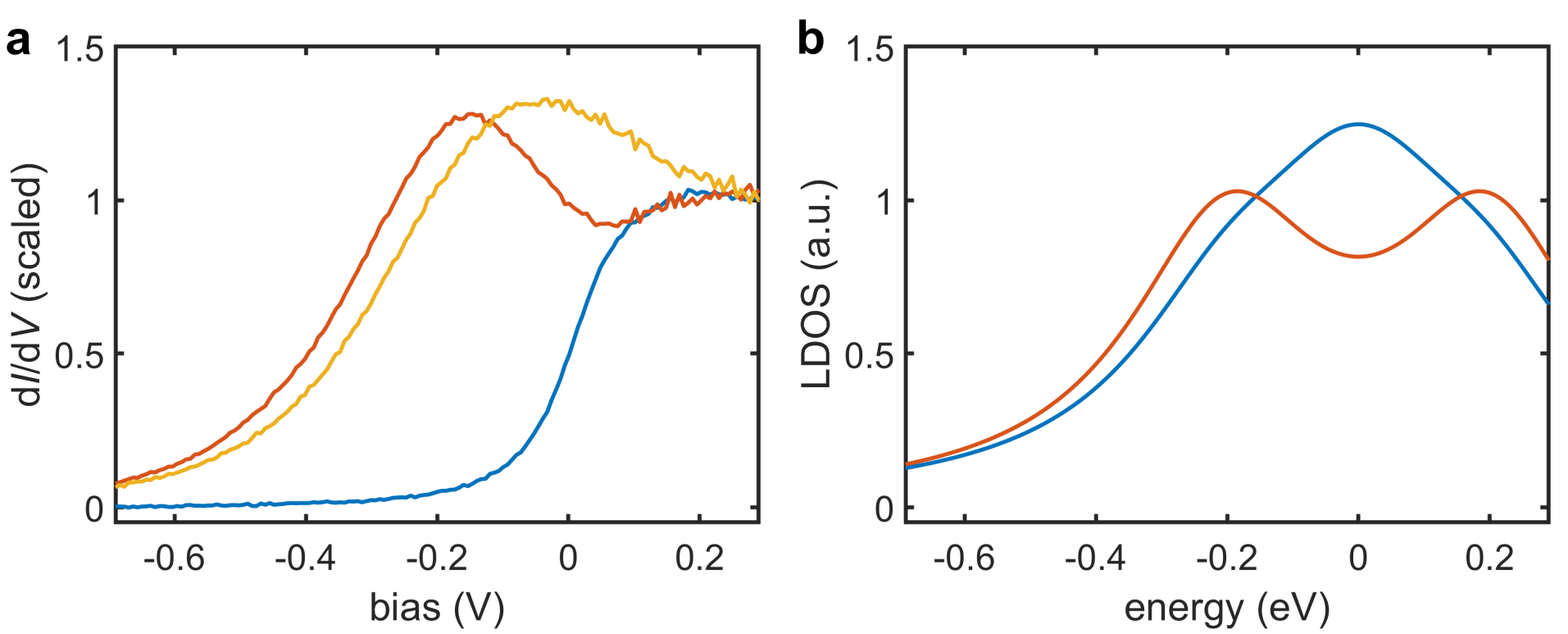}
\caption{\textbf{a,b} Spectra measured in the middle and edge sites within the unit cell of a trimer chain (panel a) and the corresponding simulated LDOS curves (panel b).}
\label{fig:S_trimer}
\end{figure}

The large energy broadening makes it difficult to extract the domain wall energies from the point spectra. In addition, the higher lying states coincide with the conduction band of the chlorine layer and hence cannot be reliably detected. This is also the case with the trimer rows as illustrated in Fig.~\mbox{\ref{fig:S_trimer}. This figure compares the spectra measured in the middle and edge sites of unit cell within the bulk of a trimer row with the background measured outside the trimer row. The bias axis has been converted to an energy axis through the measured on-site energy of 3.49 V.}

\begin{figure}[!h]
\centering
\includegraphics [width=.6\columnwidth] {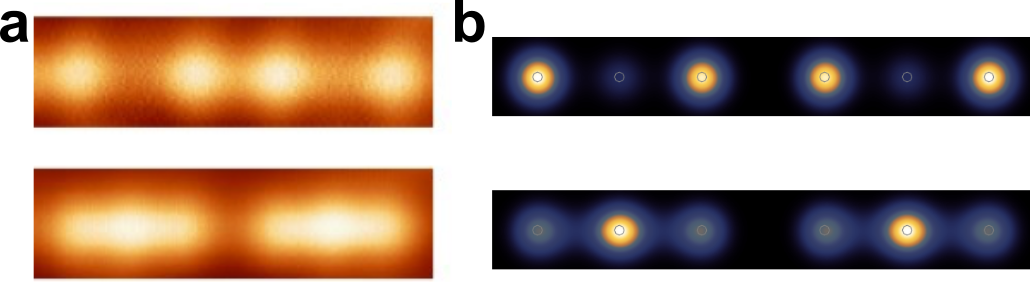}
\caption{\textbf{a,b} Comparison of the experimental (panel a) and simulated (panel b) LDOS maps at energies corresponding to the middle of the lowest band on a trimer chain (bottom panels) and at an energy corresponding to the on-site energy (top panels).}
\label{fig:S_trimer_map}
\end{figure}

It is clear in Fig.~\mbox{\ref{fig:S_trimer}} that we can see the lowest band and part of the middle band contribution in the spectra, but the on-set of the chlorine layer conduction band makes it impossible to resolve the highest band. In addition, the strong broadening (the simulated spectra have been broadened with a value that matches the measurement on a single vacancy, $0.18\pm0.01$ eV) makes it impossible to resolve the gaps between the bands. However, the contrast in the LDOS maps does still match the expected results, see Fig.~\mbox{\ref{fig:S_trimer_map}}.

\newpage
\section{Domain wall in a trimer chain consisting of two vacancies}
\begin{figure}[!h]
\centering
\includegraphics [width=.8\columnwidth] {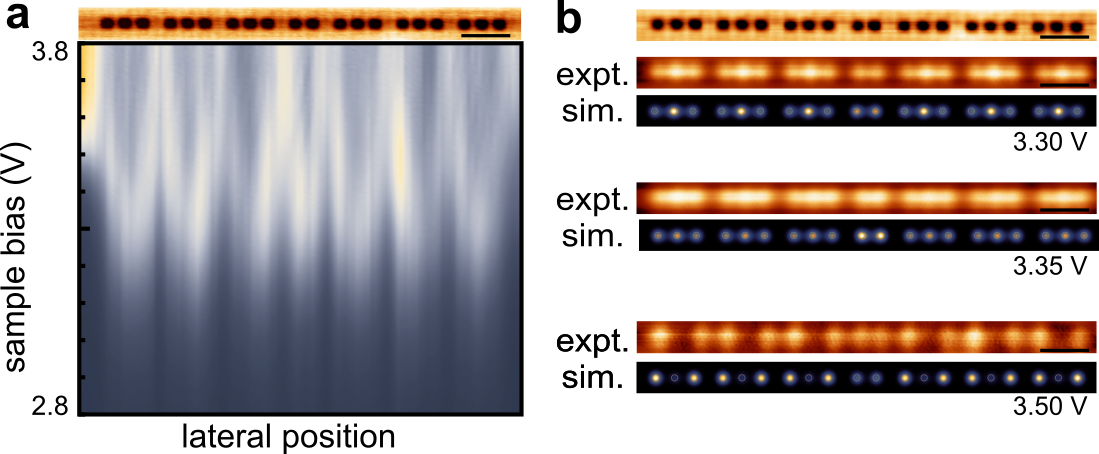}
\caption{Experimental realization of a domain wall consisting of two vacancies in a trimer chain. \textbf{a} STM topography (top) and d$I$/d$V$ spectra recorded on a line along the trimer chain. \textbf{b} Experimental d$I$/d$V$ and simulated LDOS maps at the biases indicated in the figure. Theoretical plots are based on a tight-binding model with $t_1=0.14$ eV, $t_2=t_3=0.04$ eV. Scale bars, 2 nm.}
\label{fig:s_1}
\end{figure}

In addition to the single vacancy domain wall shown in Fig.~2 in the main text, we have also fabricated a domain wall consisting of two vacancies in a trimer chain. Fig.~\ref{fig:s_1}a shows STM topography and d$I$/d$V$ spectra measured along the trimer chain. The two vacancy domain wall results in localized states that are bonding and antibonding combinations of the vacancy site wavefunctions. These states can be easily visualized by spatially-resolved d$I$/d$V$ maps, where the bonding and anti-bondings states are visible at biases around 3.35 V and 3.5 V, respectively (see Fig.~\ref{fig:s_1}b). 

\newpage
\section{Comparison between calculated and measured LDOS maps as a function of the energy for the domain wall with the weakest coupling}

\begin{figure}[!h]
    \centering
    \includegraphics[width=.8\columnwidth] {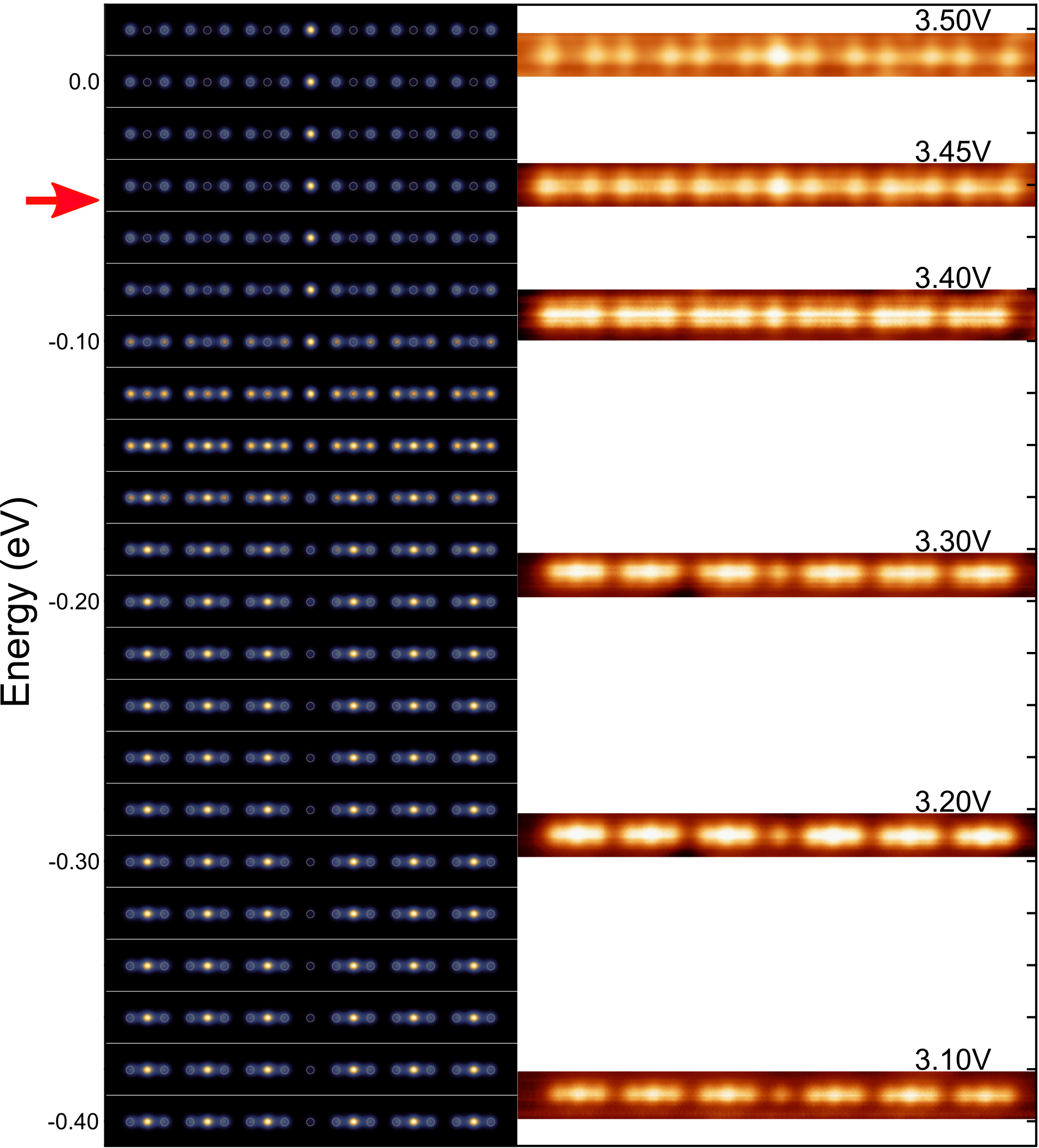}
    \caption{Comparison between calculated and measured LDOS maps as a function of the energy for the domain wall with the weakest coupling.}
    \label{fig:my_figure lable}
\end{figure}

\newpage

\section{Domain wall states in coupled dimer chains}
\begin{figure}[!h]
    \centering
    \includegraphics[width=0.9\columnwidth]{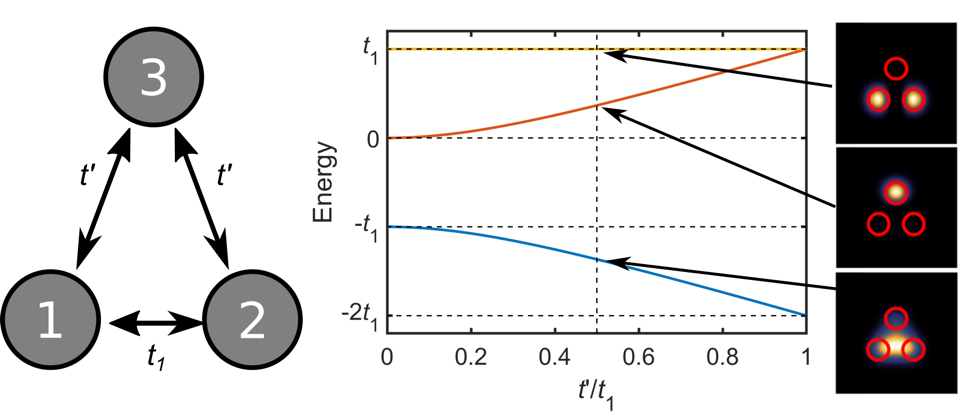}
    \caption{Schematic of a domain wall segment for the $AA\rightarrow BA$ structure and the energy levels as a function of the interchain coupling.}
    \label{fig:domain_wall_states_cd}
\end{figure}

The energies of the domain wall states are not equal in the different cases and they do depend on the values of the hoppings. To get a general idea on the behavior, we can consider states arising only from the domain wall segment. For example, Fig.~\mbox{\ref{fig:domain_wall_states_cd}} shows a schematic of the energy levels as a function of the interchain ($t'$) coupling and the simulated LDOS maps for the experimental value of the interchain coupling for the $AA\rightarrow BA$ domain wall. 

For the experimental value of the interchain coupling, the bulk spectrum has two gaps as explained in the manuscript. In this case, the highest energy domain wall state overlaps with a bulk band and hybridizes with it. Consequently, the system has two states within the gaps. Furthermore, the energies of these states are influenced on how they are coupled with the bulk chains (similarly to the case of the trimer chain). 

\begin{figure}[!h]
	\centering
	\includegraphics{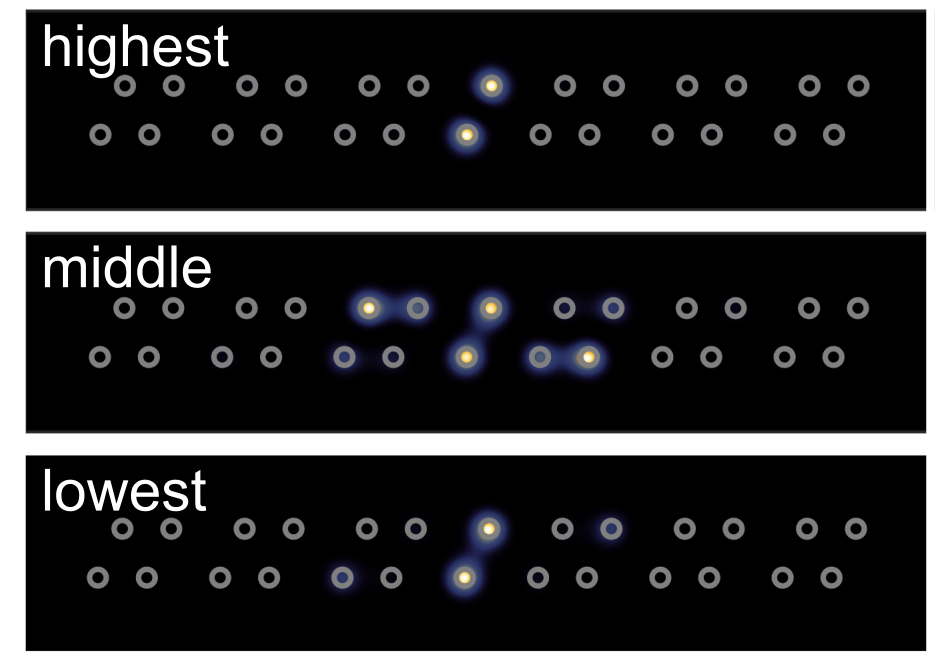}
	\caption{Plots of the three in-gap state wavefunctions for the $AA\rightarrow BB$ domain wall in a coupled dimer chain.}
	\label{fig:SI:AA_BB}
\end{figure}

In the case of the $AA\rightarrow BB$ domain wall, there are three in-gap states. The two sites forming the domain wall contribute two states that form bonding and antibonding combinations. The third in-gap state is a hybrid between the domain wall and the middle band. Fig.~\mbox{\ref{fig:SI:AA_BB}} shows the wavefunctions of these states. 

\newpage
\section{Additional results on coupled dimer chains}
\begin{figure}[h]
\centering
\includegraphics [width=.5\columnwidth] {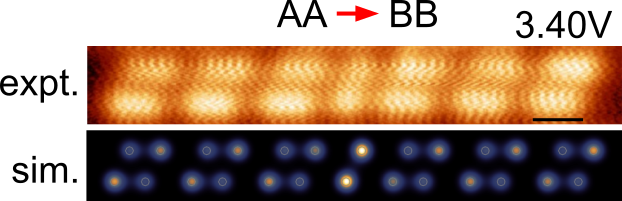}
\caption{Experimental d$I$/d$V$ and simulated LDOS maps of the third domain wall states of the $AA \rightarrow BB$ coupled dimer chain. Scale bars, 2 nm.}
\label{fig:CD}
\end{figure}

Fig.~\mbox{\ref{fig:CD}} shows experimental results for the third in-gap state on the structure $AA\rightarrow BB$.

\begin{figure}[h]
\centering
\includegraphics [width=.6\columnwidth] {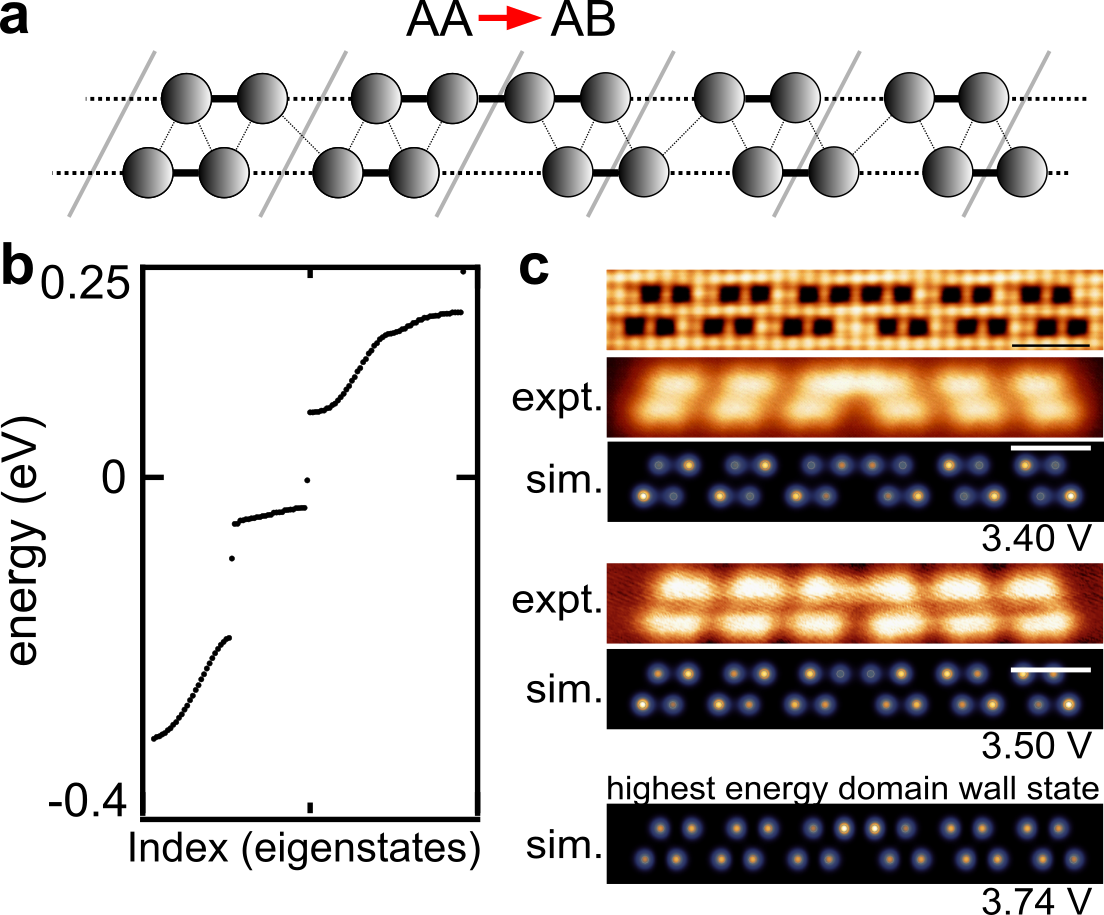}
\caption{Experimental realization of coupled dimer chain with a $AA \rightarrow AB$ domain wall. \textbf{a} Schematic of the structure. \textbf{b} The energy level diagram of the domain wall shown in panel a. \textbf{c} Experimental d$I$/d$V$ and simulated LDOS maps at the biases indicated in the figure. Scale bars, 2 nm.}
\label{fig:s_5}
\end{figure}

We have also fabricated additional coupled dimer chain with a $AA \rightarrow AB$ domain wall and characterized the domain wall states using d$I$/d$V$ spectroscopy and mapping. The calculated band structure for the domain wall shown in Fig.~\ref{fig:s_5}a is presented in Fig.~\ref{fig:s_5}b. While the  d$I$/d$V$ maps taken at low biases show delocalized states along the whole chain, the domain wall states are localized at the domain boundary and can be visualized in the spatially-resolved d$I$/d$V$ maps taken at corresponding bias voltages shown in the Fig. \ref{fig:s_5}c. Due to the experimental energy broadening, the LDOS maps show some intensity also away from the domain wall. However, the individual simulated wavefunctions of the domain wall states show that they are localized on the domain wall. Comparison of the experimental d$I$/d$V$ maps with the simulated LDOS maps shows subtle differences in the observed contrast. This is likely to be due to the onset of the conduction band of the Cl-layer (the vacancy state energy is just below the conduction band on-set). This effect is more strongly visible at higher biases, and prevents us from accessing the highest lying domain wall state (theoretically predicted to be 0.245 eV above the on-site energy).

\bibliography{bibliography}